\documentclass[prx,twocolumn,superscriptaddress,floatfix,longbibliography]{revtex4-1}

\usepackage[utf8]{inputenc}
\usepackage[T1]{fontenc}
\usepackage{graphicx, placeins}
\usepackage{color}
\usepackage{comment}
\usepackage{amsmath}

\begin{document}

\title{Fast universal two-qubit gate for neutral fermionic atoms in optical tweezers}

\author{Jonathan Nemirovsky} 
\author{Yoav Sagi}

\email[Electronic address: ]{yoavsagi@technion.ac.il}

\affiliation{Physics Department and Solid State Institute, Technion - Israel Institute of Technology, Haifa 32000, Israel}

\date{\today}
\begin{abstract}
An array of ultracold neutral atoms held in optical micro-traps is a promising platform for quantum computation. One of the major bottlenecks of this platform is the weak coupling strength between adjacent atoms, which limits the speed of two-qubit gates. Here, we present a method to perform a  fast universal $\sqrt{\text{SWAP}}$ gate with fermionic atoms. The basic idea of the gate is to release the atoms into a harmonic potential positioned in between the two atoms. By properly tailoring the interaction parameter, the collision process between the atoms generates entanglement and yields the desired gate. We prove analytically that in the limit of broad atomic wave-packets, the fidelity of the gate approaches unity. We demonstrate numerically that with typical experimental parameters, our gate can operate on a microsecond timescale and achieves a fidelity higher than $0.998$. Moreover, the gate duration is independent of the initial distance between the atoms. A gate with such features is an important milestone towards all-to-all connectivity and fault tolerance in quantum computation with neutral atoms.
\end{abstract}

\maketitle
\section{Introduction}

Quantum mechanics poses a computational challenge: the dimension of the Hilbert space grows exponentially with the system size. As a result, a classical simulation of a many-body quantum system quickly becomes intractable as the number of particles increases. The solution to this problem, as first pointed out by Richard Feynman \cite{feynman1982simulating}, is to use a quantum computational machine (``quantum computer'') instead of a classical one \cite{Lloyd1996}. In addition to efficient simulation of quantum systems, a quantum computer will allow polynomial solutions to complex mathematical problems such as factoring and searching \cite{1107002176,Ladd2010}. The effort to build a quantum computer is ongoing for more than 25 years \cite{divincenzo1995quantum}. Many physical systems have been suggested as carriers of quantum information, including superconducting circuits \cite{Devoret2013,Wendin2017,Gu2017,Arute2019,Gong2019}, trapped ions \cite{Cirac1995,Leibfried2003,Blatt2008,HAFFNER2008,Blatt2012,Monroe2013,Bruzewicz2019,Wright2019}, ultracold atoms \cite{Bloch2008,Saffman2016,Gross2017,Weiss2017,Bernien2017}, photons \cite{Knill2001,Prevedel2007,Kok2007,Flamini2018}, defects in solids \cite{Dutt2007,Weber2010,Childress2013}, and quantum dots \cite{Imamoglu1999,Petta2005,Hanson2007,Schwartz2016}. The prevalent paradigm for quantum computation starts with initialization of the quantum bits (``qubits''), application of a series of one and two qubit gates from a small set of universal gates, and finally, a measurement of the qubits final state \cite{Divincenzo2005}. It is essential that the fidelity of the gates is high enough to achieve fault tolerance through quantum error correction \cite{1107002176}. There are advantages and disadvantages to each platform in different aspects, but at this point of time, none is fully scalable.    

A promising approach to employing ultracold neutral atoms for quantum computation is based on holding them one by one in far-off-resonance optical micro-traps (``optical tweezers'') \cite{Dumke2002,Bergamini2004,Nogrette2014,Barredo2016,Endres2016,Norcia2018,Cooper2018,Kumar2018,Barredo2018,Levine2019,Mello2019}. These experiments are performed in ultra high vacuum chambers, where the atoms can be very efficiently isolated from the environment. The optical tweezers' parameters, such as position, width and depth, can be dynamically modified by controlling the electro-optical devices that generate the beams. Quantum information is usually encoded in internal states of the atoms. Two-qubit gates exploit the interaction between the atoms, whose range can be very short, in the case of van der Waals interaction, or considerably longer, in the case of dipole-dipole interaction. The strength of the interaction can be tuned via a Feshbach resonance \cite{Chin2010}, in the former case, or by controlling the angle or distance between the atoms, in the latter case \cite{Lahaye2009,Leseleuc2019}. A large dipole moment exists for specific ground-state atoms \cite{Lahaye2009}, molecules \cite{Anderegg2019}, and atoms excited to a large principal quantum number (Rydberg atoms) \cite{Saffman2016,Labuhn2016,Leseleuc2019,Omran2019}. The current maximum fidelity of a two qubit gate with Rydberg atoms is around $0.97$ \cite{Levine2019}. However, Rydberg atoms have a relatively short lifetime and they are sensitive to stray electric fields.

An alternative approach is to work with ground state atoms, prepared in the lowest vibrational state of the tweezers \cite{Serwane2011,Kaufman2012}. A universal two-qubit $\sqrt{\text{SWAP}}$ gate can be implemented by allowing the atoms to tunnel between the traps and exploiting the short-range interaction  \cite{hayes2007quantum}. The operation of the gate is based on exchange blockade manifested through the symmetry of the two particles wave-function and the onsite interaction. The duration of the gate depends on the tunneling rate, which in turn is set by the distance between the tweezers. By moving the traps closer or further away, it is possible to effectively switch ``on'' or ``off'' the tunneling.  To maintain high fidelity for the gate, this movement should not excite the atoms. Most naturally, this is accomplished by following an adiabatic motion \cite{hayes2007quantum}. However, this implies a relatively slow gate, which eventually compromises the overall fidelity of the whole calculation. 

Here, we present a novel approach to perform a fast and robust universal $\sqrt{\text{SWAP}}$ gate with ground state atoms that interact through a tunable contact-like potential. The basic idea of our gate is to turn off the tweezers and turn on an auxiliary harmonic potential centred midway between the two atoms. In this potential, the atomic wave packets can be described as squeezed coherent states. By tailoring the s-wave scattering length, the scattering process between the atoms gives rise to the desired relative phase shift between the even and odd components composing the two-particle wave-function. The gate duration is set by the harmonic trap period. In principle, it can be as short as the experimental resources allow. We demonstrate numerically that using realistic parameters, a gate operating at a fidelity of $0.998$ can be achieved in approximately $20\mu$s. In fact, the gate duration can be further reduced, limited only by the available optical power in the beam that generates the central harmonic potential. We also show that using shortcut to adiabaticity driving, it is possible to reach the same level of performance with a time-dependent harmonic potential. Our approach is general and can easily be applied in current experiments. Importantly, it allows for more than $10^5$ gate operations during an experimentally achievable coherence time of atoms in a tweezers array \cite{Norcia2019}.

The structure of this paper is as follows: in section \ref{sec_adiabatic_gate} we define the model and provide a general solution for the adiabatic gate. This solution establishes the linear scaling between the tunneling time and the adiabatic gate duration. In section \ref{sec_fast_gate} we analyze the new gate. We analytically solve the the two-particle dynamics and prove that the fidelity of the gate approaches unity as the squeezing parameter of the atomic wave-packets in the central harmonic trap increases. We demonstrate numerically the operation of the gate with realistic experimental parameters in section \ref{sec_numerical_simulations}. In section \ref{sec_STA} we introduce a scale-invariant driving of the harmonic trap to achieve two goals: 1. allow for a continuous initiation and termination of the harmonic trap. 2. further improve the fidelity by increasing the squeezing parameter. We conclude in section \ref{sec_discussion}. 

\section{An adiabatic $\sqrt{\text{SWAP}}$ gate}\label{sec_adiabatic_gate}
Before explaining the fast $\sqrt{\text{SWAP}}$ gate, it is instructive to examine first the adiabatic one. We consider two tightly focused Gaussian optical traps whose parameters, such as position or trap depth, can be dynamically controlled. A single fermionic atom is prepared in the ground vibrational state of each trap. There are several internal states to each atom, but we restrict ourselves here to two that constitute the qubit states, denoted by $|\downarrow\rangle$ and $|\uparrow\rangle$. The distance between the traps is $d$.

In the adiabatic gate, the tweezers are always present. Hence, we employ a tight-binding approximation and write the Hamiltonian as: $\hat{H}=J\left(\hat{u}^\dagger_1 \hat{u}_2+\hat{u}^\dagger_2 \hat{u}_1+\hat{d}^\dagger_1 \hat{d}_2+\hat{d}^\dagger_2 \hat{d}_1\right)+U\left(2\hat{u}^\dagger_1 \hat{u}_1 \hat{d}^\dagger_1 \hat{d}_1+2\hat{u}^\dagger_2 \hat{u}_2 \hat{d}^\dagger_2 \hat{d}_2\right)$, where $J$ is the tunneling energy, $U$ is the on-site particle-particle interaction energy, and $\hat{u}^\dagger_i$ ($\hat{d}^\dagger_i$) is the fermionic creation operator for a particle in trap $i$ at a state $|\uparrow\rangle$ ($|\downarrow\rangle$). The $\sqrt{\text{SWAP}}$ gate unitary operator is diagonal in the basis of the singlet,  $\frac{1}{\sqrt{2}}(\hat{u}^\dagger_1\hat{d}^\dagger_2-\hat{u}^\dagger_2\hat{d}^\dagger_1)|\text{vac}\rangle$, and triplet states, $\{\frac{1}{\sqrt{2}}(\hat{u}^\dagger_1\hat{d}^\dagger_2+\hat{u}^\dagger_2\hat{d}^\dagger_1)|\text{vac}\rangle,\hat{u}^\dagger_1\hat{u}^\dagger_2|\text{vac}\rangle,\hat{d}^\dagger_1\hat{d}^\dagger_2|\text{vac}\rangle \}$, with eigenvalues $e^{i\pi/2}$ and $1$, respectively. For a given tunneling rate, $J_g$, a proper choice of the interaction $U_g=\frac{2}{\sqrt{3}}J_g$ yields the required gate after a time $t_\text{gate}=\frac{\pi\sqrt{3}}{4}\hbar J^{-1}_g$. Owing to symmetry, only the singlet state evolves under $\hat{H}$ into a state with double occupancy in each of the traps, which acquires due to the on-site interaction term an additional phase relative to the triplet states. 

In the adiabatic gate, the traps are initially far apart such that $J$ is essentially zero, then slowly brought closer to initiate the tunneling, and finally are separated again to stop the gate. We simulated numerically such a motion with two fermionic ${}^6\text{Li}$ atoms (see also section \ref{sec_numerical_simulations}). We use realistic experimental parameters -- the two Gaussian optical potentials, $-V_0e^{-\frac{2 x^2}{ \sigma^2}}$, have a waist $\sigma=700$nm and a depth of $V_0=20.38\mu \text{K}\times k_B$, where $k_B$ is the Boltzmann constant. The traps are initially separated by $d(t=0)=2.329\mu$m, a distance at which the tunneling is completely negligible. Using a smooth cosine curve for $d(t)$, we find that in order to achieve a fidelity higher than $0.99$, the duration of the gate should be longer than $\sim 320\mu$s. This result is consistent with $t_\text{gate}=\frac{\pi\sqrt{3}}{4}\hbar \bar{J}^{-1}_g=284.9 {\mu}$s, which we calculate using the stationary solution given above and using the time-averaged value along the motion $\bar{J_g}=\frac{1}{t_\text{gate}}\int_0^{t_ \text{gate}}J(t')dt'$. As we show below, this adiabatic time is more than an order of magnitude larger than what can be achieved using our fast gate.

\section{Fast $\sqrt{\text{SWAP}}$ gate}\label{sec_fast_gate}
\begin{figure}[ht]
	\centering
	\includegraphics[width=1.0\linewidth]{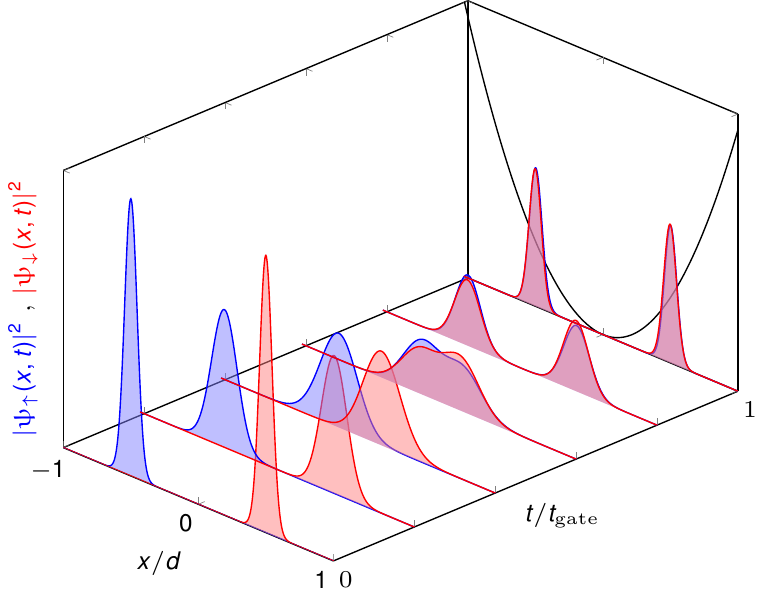}
	\caption{\textbf{Fast $\sqrt{\text{SWAP}}$ gate.} At $t<0$ two atoms are in two separate optical micro-traps. We depict the single particle probability distributions of each atom. The two atoms with spins $\uparrow$ (blue) and $\downarrow$ (red) are initially centered around $x=-\frac{d}{2}=-1.165\mu$m and $x=\frac{d}{2}=1.165\mu$m, respectively. At $t=0$ the micro-traps are shut off and a harmonic trap is turned on (black line). The atomic wave-packets start to expand and move towards the center of the trap, where they collide at around $t_\text{gate}/2$. The interaction during the scattering process entangles the atoms. Hence, at the end time, each atom is split between the two traps. The gate ends after half of the harmonic trap period $t_\text{gate}=\pi/\omega_0$, which is $21.84\mu \text{s}$ in this calculation. Then, the harmonic trap shuts off and the micro-traps are turned on.}
	\label{fig:fast_gate_operation}
\end{figure}

In Fig. \ref{fig:fast_gate_operation} we plot the atomic wave-packets probability distributions during our fast gate. To shorten the gate time, we forgo the requirement that each atom will be localized in one of the traps during the operation of the gate. Our gate starts at $t=0$ by switching off the two micro-traps and concurrently switching on a harmonic trap centred midway between the traps. After half the harmonic trap period, $\pi/\omega$, the gate ends by switching off the harmonic trap and turning on the two micro-traps at their original location. Without interactions between the atoms, this realizes a SWAP gate, which by itself is a useful building block in quantum computation platform. With interactions, however, the two atoms scatter on each other as they collide in the harmonic trap. With a proper choice of the interaction strength, a $\pi/2$ phase shift develops between the even and odd components of the two-body wave-function. This transforms the gate into an entangling $\sqrt{\text{SWAP}}$ gate.

Leaving the tight-binding approximation, the two-particle system is  described by the Hamiltonian \cite{giorgini2008theory}: 
\begin{equation}
\label{HamiltonianFermions}
\begin{split} %\in \{\uparrow ,\downarrow \}
\hat{H}(t)= & \sum\limits_{s\in \{\uparrow ,\downarrow \}}{\int{d\mathbf{x}\,}\hat{a}_{s}^{\dagger }(\mathbf{x})\left[ -\frac{{{\hbar }^{2}}}{2m}\nabla _{\mathbf{x}}^{2}+V_{\text{ext}}( \mathbf{x},t)\right]}\hat{a}_{s}(\mathbf{x})\,\,+ \\
&
+\iint{d\mathbf{x}}d{{\mathbf{x}}^{\prime }}V_{\text{int}} (\mathbf{x}-{{\mathbf{x}}^{\prime }})\hat{a}_{\uparrow }^\dagger({\mathbf x}) \hat{a}_{\downarrow }^\dagger({\mathbf x}^{\prime })\hat{a}_{\downarrow }({\mathbf x}^{\prime })\hat{a}_{\uparrow }({\mathbf x}) \, \, \, , 
\end{split}
\end{equation}
where the index $s\in\{\uparrow,\downarrow\}$ denotes the spin state, the operators $\hat{a}_{s}^{\dagger }$ and $\hat{a}_{s}$ are creation and annihilation operators, satisfying the fermionic anticommutation relations $\{\hat{a}_s(\mathbf{x}),\hat{a}^\dagger_{s'}(\mathbf{x})\}=\delta _{s}^{s'}\delta (\mathbf{x}-\mathbf{x}')$, with $\delta _{s}^{s'}$ and $\delta(x)$ being the Kronecker and Dirac delta functions, respectively. The single-body external potential, $V_{ext}( \mathbf{x},t)$, is initially composed of the two Gaussian optical traps, but at $t=0$ they are replaced with the harmonic potential $\frac{1}{2}k_0 x^2$.  The short-range contact interaction can be approximated as $V_{\text{int}} (\mathbf{x}-{{\mathbf{x}}^{\prime }})=F(B)\delta_{\text{reg.}}(|\mathbf{x}-\mathbf{x}^{\prime}|)$, where $\delta_{\text{reg.}}$ is a regularized delta-like potential \cite{giorgini2008theory}. The coupling constant, $F(B)$, can be tuned via the magnetic field, $B$, near an s-wave Feshbach resonance \cite{Chin2010}. At low temperatures, scattering to higher partial waves is negligible. Moreover, due to the fermionic symmetry, only atoms with opposite spins interact through the s-wave scattering process.

The two-particle wave-function can be written as 
\begin{align}
&|\psi(t)\rangle=\\&\sum\limits_{s,s'\in \{\uparrow ,\downarrow \}}\int d\mathbf{x}_1d\mathbf{x}_2 \psi_{ss'}(\mathbf{x}_1,\mathbf{x}_2,t)\hat{a}_s^\dagger(\mathbf{x}_1)\hat{a}_{s'}^\dagger(\mathbf{x}_2)|\text{vac}\rangle \ \ ,\nonumber
\end{align}
where $|\text{vac}\rangle$ is the vacuum state with no atoms. The dynamics is given by the two-particle Schr\"odinger equation
\begin{eqnarray}
i\hbar \frac{\partial {{\psi }_{ss' }}}{\partial t}&=\big[
-\frac{{{\hbar }^{2}}}{2m}\left( \nabla _{x_1}^{2}+\nabla _{x_2}^{2} \right)+
\frac{1}{2}k_0\left({x_1}^2+{x_2}^2\right)+\nonumber\\\label{Eq_schroedinger_1}
&+(1-\delta_s^{s'})F(B)\delta_{\text{reg}} (x_1-x_2) \big] {{\psi }_{ss'}} .
\end{eqnarray}
Changing coordinates to $X=\frac{{{x}_1}+{{x}_2}}{\sqrt{2}}$ and $x=\frac{{{x}_{1 }}-{{x}_2}}{\sqrt{2}}$ \cite{busch1998two}, this equation becomes 
\begin{eqnarray}
i\hbar \frac{\partial {{\psi }_{ss' }}}{\partial t}&=\big[ -\frac{{{\hbar }^{2}}}{2m}\left( \nabla _{X}^{2}+\nabla _{x}^{2} \right)+ \frac{1}{2}k_0\left(X^2+x^2\right)+\nonumber\\
&+(1-\delta_s^{s'})\gamma\delta_{\text{reg}} (x) \big] {{\psi }_{ss'}}.
\end{eqnarray}
with $\gamma = \frac{1}{\sqrt{2}}F(B)$.

The initial condition of the gate is one atom in each Gaussian trap. If the traps are far enough, the initial state is approximately a product state of the form $\psi _{s,s'}({{x}_1},{{x}_2},0)=\varphi(x_1-\frac{d}{2})\varphi(x_2+\frac{d}{2})$,
where  $\varphi(x)=\frac{1}{\sqrt{2\pi }{{w}_{0}}}\exp \left( -\frac{x^2}{2w_0^2} \right)$. This initial condition remains a product state also in the $x,X$ coordinates: $\psi_{s,s'} (x,X,0)=\varphi(x-d/\sqrt{2})\varphi(X)$. Since the Hamiltonian is a sum of two commuting operators, operating separately on $x$ and $X$, the solution at all times remains separable in the coordinates $x,X$ - i.e., $\psi_{s,s'} (x,X,t)=\psi_0 (X,t)\psi_1 (x,t)$. The solution is determined by the equations
\begin{equation}\label{harmonic}
i\hbar \frac{\partial }{\partial t}{{\psi }_{0}}(X,t)=\left[ -\frac{{{\hbar }^{2}}}{2m}\nabla _{X}^{2}+\frac{1}{2}{{k}_{0}}{{X}^{2}} \right]{{\psi }_{0}}(X,t)
\end{equation}
and
\begin{multline}
\label{contact_eq}
i\hbar \frac{\partial }{\partial t}{\psi_1}(x,t)= \Big[ -\frac{{{\hbar }^{2}}}{2m}\nabla _{x}^{2}+\frac{1}{2}{{k}_{0}}{{x}^{2}}\\+(1-\delta_s^{s'})\gamma{{\delta }_{\text{reg}}}(x) \Big]{\psi_1}(x,t) \ \ .
\end{multline} 

The decomposition of the solution to two independent equations in lower dimensions greatly simplifies the analysis of the gate. Furthermore, at the end of the gate, at ${t_{\text{gate}}=\pi/\omega_0}$ with $\omega_0=\sqrt{\frac{k_0}{m}}$, the wave-function $\psi_0(X,t_{\text{gate}})$ is identical to $\psi_0(X,t=0)$ up to a global phase, independent of the atomic spins. Thus, the key to achieve a $\sqrt{\text{SWAP}}$ gate is that
$\psi_1(x,0)=e^{i\phi_{s,s'}}\psi_1(x,t_{\text{gate}})$ with $\phi_{\uparrow\downarrow}+\pi/2=\phi_{\uparrow\uparrow}=\phi_{\downarrow\downarrow}$, namely that the wave-function $\psi_1$ will return to itself at the end of the gate, up to a constant global phase, plus a phase of $\pi/2$ in the case of opposite spins. As we show below, this can done with a proper choice of $F(B)$.

For $\gamma=0$, the solutions of Eq. (\ref{contact_eq}) are displaced squeezed coherent states \cite{jadczykcomment,marhic1978oscillating,kim2003time,kryuchkov2013minimum,a2019time}:
\begin{equation}
\label{wave_packet_expr}
{{\varphi }_{\pm }}(x,t)={{\left( \frac{A(t)}{\sqrt{\pi }} \right)}^{1/2}}{{e}^{-\frac{i}{2}\Theta (t)}}{{e}^{-\frac{{{\left| x\pm {{x}_{c}}(t) \right|}^{2}}}{w(t)^2}}}{{e}^{\pm i{{p}_{c}}(t)x/\hbar }}
\end{equation}with,
\begin{align}
A(t)&=\sqrt{\frac{m\omega_0 }{\hbar }}\frac{1}{\sqrt{\cosh (2r)+\sinh (2r)\cos (2\omega_0 t)}}\nonumber\\
w(t)^2&=\frac{2\hbar }{m\omega_0 }\left[ \frac{1+\tanh (r){{e}^{-2i\omega_0 t}}}{1-\tanh (r){{e}^{-2i\omega_0 t}}} \right]\nonumber\\
{{e}^{-i\Theta (t)}}&={{\left[ \frac{1+\tanh (r){{e}^{+2i\omega_0 t}}}{1+\tanh (r){{e}^{-2i\omega_0 t}}} \right]}^{1/2}}{{e}^{-i\omega_0 t}}\nonumber\\
{{x}_{c}}(t)&=\frac{d}{\sqrt{2}}\cos (\omega_0 t)\nonumber\\
{{p}_{c}}(t)&=m{{\dot{x}}_{c}}(t)=-\frac{m\omega_0 d}{\sqrt{2}}\sin (\omega_0 t) \ \ .\label{wave_packet_expr_param}
\end{align}
%which are at $\pm\frac{d}{\sqrt{2}}$ at $t=0$. 
The squeezing parameter, $r$, is determined by the initial condition $w(0)=w_0$,
\begin{equation}\label{Eq_squeezing_parameter}
r=\ln\left[\frac{w_0}{\sqrt{2}\sqrt{\hbar/m\omega_0}}\right] \ \ .
\end{equation}
Similarly, the solution of $\psi_0(X,t)$ is a squeezed wave-packet starting at the origin with no initial momentum, satisfying $x_c(t)=0$ and $p_c(t)=0$.

The challenge now is to solve for $\psi_1(x,t)$ in the interacting case, $\gamma(t)\neq 0$. The displaced squeezed wave-packets ${{\varphi }_\pm}(x,t)$ collide with a delta-potential at the origin. Thus, the solution to Eq.(\ref{contact_eq}) can be written as a scattering wave-function, 
\begin{equation}
\label{psi_1_def}
\begin{split}
\psi_1(x,t)=\left\{ \begin{matrix} &
{{\varphi }_{+}}(x,t)+R{{\varphi }_{-}}(x,t)\,\,\,\,\,\,& \text{for}\,\,\,x>0 \, \, \, \\
& T{{\varphi }_{+}}(x,t)\,\,\,\,\,\, & \text{for}\,\,\,x<0 \, .  \\
\end{matrix} \right.
\end{split}
\end{equation}
The coefficients $R$ and $T$ are found by imposing the necessary boundary conditions at $x=0$:
\begin{enumerate}
	\item Continuity 
	\begin{equation}\label{Eq_boundary_condition_1}
	{{\varphi }_{+}}(0,t)+R{{\varphi }_{-}}(0,t)=T{{\varphi }_{+}}(0,t)
	\end{equation}
	\item Momentum conservation 
	\begin{equation}\label{Eq_boundary_condition_2}
	\left. -\frac{{{\hbar }^{2}}}{2m}\frac{\partial {\psi_1}}{\partial x}(x,t) \right|_{x=0^- }^{x=0^+ }=\gamma(t) {\psi_1}(0,t) ,
	\end{equation}
\end{enumerate}

\begin{figure*}[ht!]
	\centering
	\includegraphics[width=1.0\linewidth]{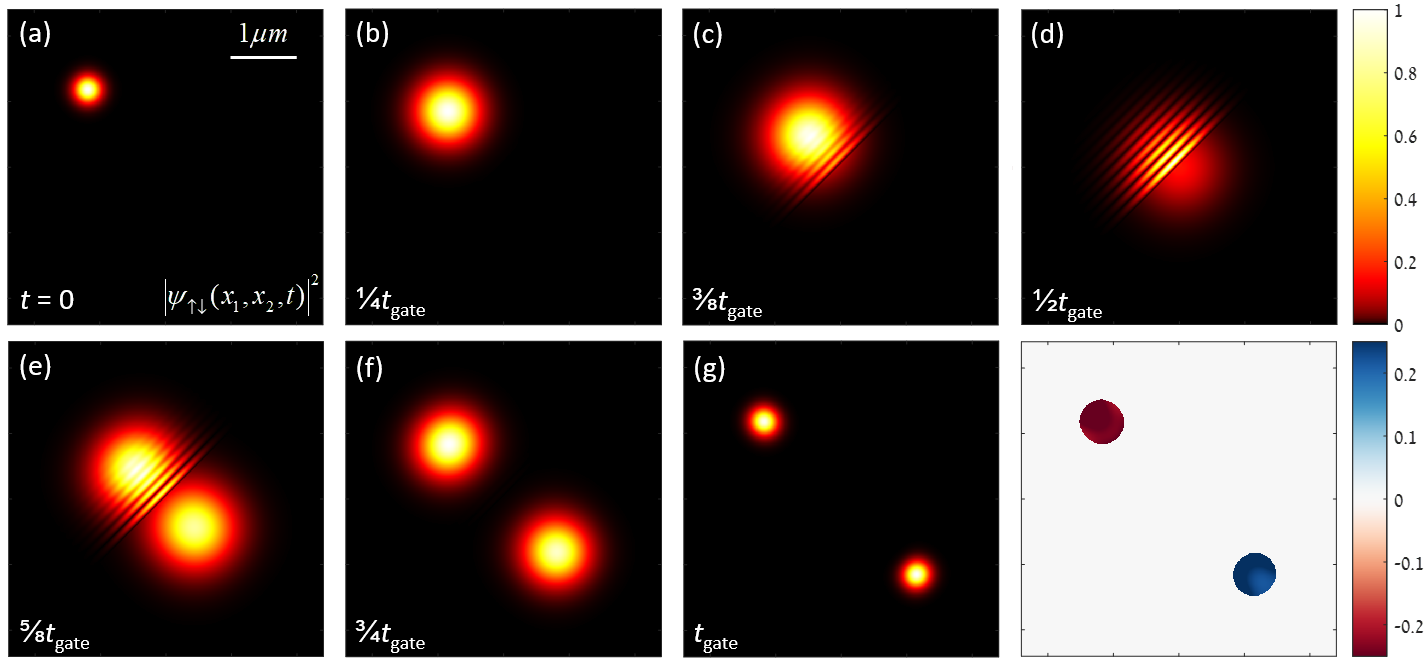}
	\caption{\textbf{Numerical simulation of a fast $\sqrt{\text{SWAP}}$ gate between two atoms with opposite spins.} Panels (a)-(g) depict the two-particle probability distribution, $|\psi_{\uparrow \downarrow}(x_1,x_2,t)|^2$, for different times during the operation of the gate, with $x_1$ and $x_2$ being the horizontal and vertical axes, respectively. The gate duration is $t_\text{gate}=21.84\mu$s. \textbf{(a)} Initially, the two particle wave-function is the product of the two atoms' ground state solutions (one centered around $x_1=-1.1645\mu$m and the other centered at $x_2=1.1645\mu$m). \textbf{(b)} As time progresses, the distribution initially  broadens and shifts towards the center -- a direct manifestation of the displaced squeezed wave-packet dynamics given in Eq.(\ref{wave_packet_expr}). \textbf{(c)-(d)} The fringes that gradually develop are the result of the interference between the incident and reflected wave-packets of Eq.(\ref{psi_1_def}). The symmetry of the interference pattern reflects the fact that the scattering from the delta-like potential occurs only in the $x=\frac{x_1-x_2}{\sqrt{2}}$ coordinate - see Eq.(\ref{contact_eq}). \textbf{(f)} Towards the end of the gate, the wave-packets return to their initial size and location.\textbf{(g)} The two-particle distribution at the final time is clearly entangled. \textbf{(h)} The phase of the wave-function at $t_\text{gate}$, with colorbar in units of $\pi$ radians. The relative $\pi/2$ phase between the two wave-packets shows that indeed it is a $\sqrt{\text{SWAP}}$ gate.}
	\label{fig:two_particle_simulation}
\end{figure*} 

From the continuity condition and using Eq.(\ref{wave_packet_expr}), we obtain $1+R=T$. To realize a $\sqrt{SWAP}$ gate, we require $R/T=\pm i$, which yields $T=\frac{1}{\sqrt{2}}\exp(\pm i \pi/4)$. Using Eq.(\ref{wave_packet_expr}), we find the $x$-derivative of the solution, 
\begin{equation}
\frac{\partial }{\partial x}{{\varphi }_{\pm }}(x,t)=\left[ \pm \frac{i}{\hbar }{{p}_{c}}(t)-2\frac{x\pm {{x}_{c}}(t)}{w{{(t)}^{2}}} \right]{{\varphi }_{\pm }}(x,t).
\end{equation}
Finally, we use Eq.(\ref{Eq_boundary_condition_2}) to find $\gamma(t)$,
\begin{align}\label{Eq_gamma_t_1}
&\gamma (t)=\pm i\frac{{{\hbar }^{2}}}{m}\left[ \frac{i}{\hbar }{{p}_{c}}(t)-2\frac{{{x}_{c}}(t)}{w{{(t)}^{2}}} \right]= \\ 
&= \mp\frac{\omega_0 \hbar d}{\sqrt{2}}\left[ \sin (\omega_0 t)-i\cos (\omega_0 t)\frac{{{e}^{i\omega_0 t}}-\tanh (r){{e}^{-i\omega_0 t}}}{{{e}^{i\omega_0 t}}+\tanh (r){{e}^{-i\omega_0 t}}} \right] \ \ .\nonumber
\end{align}
Since $\gamma(t)$ must be a real number, it follows that we must choose a squeezing parameter such that $\tanh(r)=1$. Eq.(\ref{Eq_gamma_t_1}) then reads
\begin{equation}\label{Eq_gamma_t_2}
\gamma (t)=\mp\sqrt{2}\omega_0 \hbar d\sin(\omega_0 t) \ \ .
\end{equation} 
We therefore reach the conclusion that for $\tanh(r)=1$, tuning the interactions according to Eq.(\ref{Eq_gamma_t_2}) (i.e., by changing the applied magnetic field) yields a $\sqrt{\text{SWAP}}$ gate with a perfect fidelity. However, $\tanh(r)=1$ is nonphysical, since it requires an initial Gaussian wave-packet with an infinite width. Nonetheless, as shown below, the fidelity increases rapidly towards unity as $\tanh(r)$ increases towards one. Since the squeezing parameter is set by the ratio of the initial width of the Gaussian wave-packet to the oscillator length (see Eq. \ref{Eq_squeezing_parameter}), increasing the trapping frequency of the harmonic potential both shortens the gate duration and increases its fidelity.

Several pragmatic comments are in place at this point. First, a true harmonic trap is unbounded and therefore nonphysical. However, in experiments it can be approximated by a Gaussian potential, $-V_0 e^{-2x^2/\sigma^2}$, where near its minimum, the effective harmonic frequency is given by $\omega_0=\sqrt{\frac{4V_0}{m \sigma^2}}$ \cite{Grimm2000}. To be a good approximation, we have to require that this Gaussian potential will be broader than the distance between the traps, preferably satisfying $\sigma\gg d$. To achieve the highest fidelity, we want to increase $\omega_0$ and therefore deepen the Gaussian trap. Hence, the physical resources (i.e., available laser power) set a limit for the gate fidelity.

Second, the gate time is determined by the period of the harmonic trap and is independent of the interaction. This simplifies considerably the optimization of the gate. A scan of a single parameter, the prefactor in Eq.(\ref{Eq_gamma_t_2}), is enough to optimize the gate with a finite $r$. Moreover, the significant part of the collision between the atomic wave-packets happens over a short time interval of the order
$\frac{w(t_\text{gate}/2)}{\dot{x}_c(t_\text{gate}/2)}=\frac{2\sqrt{\hbar}}{dm^{1/2}\omega_0^{3/2}}\sqrt{\frac{1-\tanh(r)}{1+\tanh(r)}}$. Since the value of $\gamma(t)$ is most important during this interval, we find numerically that the gate fidelities achieved with a constant interaction parameter are only slightly smaller than those obtained with a time-dependent one.  

\section{Numerical simulations}\label{sec_numerical_simulations}
To demonstrate our gate and study its performance in realistic conditions, we solve numerically the time-dependent two-particle Schr\"odinger equation (\ref{Eq_schroedinger_1}) using the Beam Propagation Method (BPM) \cite{thylen1983beam}. BPM is an efficient numerical method that utilizes an operator splitting; at each time step the evolution of the wave-function due to the kinetic term is computed in Fourier-space, and then the evolution due to the time-dependent potential term is computed in real-space. Unless written otherwise, the simulations were done on a square grid of size $4.8\mu\text{m}\times 4.8\mu\text{m}$, with $576$ divisions in each direction. With this choice, the two-particle wave function practically vanishes outside the grid, and the numerical accuracy is approximately $2\cdot 10^{-3}$.

The simulations are done with the same initial conditions as in the adiabatic case (see section \ref{sec_adiabatic_gate}), namely $d(t=0)=2.329\mu$m, the waist and depth of the tweezers are $700$nm and $20.38\mu\text{K}\times\text{k}_B$, respectively. The Gaussian potential that approximates the central harmonic trap has a waist of $\sigma=11.857\mu$m, considerably larger than the distance between the atomic wave-packets. Its depth is chosen to be $U/\text{k}_B\approx 525\mu\text{K}$, such that the harmonic oscillation frequency near the center of the Gaussian potential is $\omega_0= 2\pi\times 22898$Hz. This yields a gate time of $t_\text{gate}\approx 21.84\mu$s, which is a factor 15 faster than the adiabatic gate with the same initial conditions. The gate time can be farther reduced if $U$ is increased, limited only by the available laser power. All the parameters used in the simulations are attainable in the current generation of experiments.
%$U=7.2483\cdot 10^{-27}$J

The result of a simulation starting with two atoms with opposite spins is shown in Fig. \ref{fig:two_particle_simulation}. Panels (a)-(g) depict the two-particle wave-function probability distribution, $|\psi_{\uparrow \downarrow}(x_1,x_2,t)|^2$, for different times during the gate's operation, with $x_1$ and $x_2$ being the horizontal and vertical axes, respectively. Initially, the two particle wave-function is the product of the two atoms' ground state solutions (one centered around $x_1=-1.1645\mu$m and the other centered at $x_2=1.1645\mu$m). In panel (b), which corresponds to $t_\text{gate}/4$, the distribution shifts towards the center of the large Gaussian potential and broadens -- a direct manifestation of the displaced squeezed wave-packet dynamics given in Eq.(\ref{wave_packet_expr}). Panel (c) already shows a clear sign of the delta-potential scattering process, as described by Eq.(\ref{psi_1_def}). The fringes are the result of the interference between the incident and reflected wave-packets. The transmitted wave-packets do not exhibit similar interference patterns, as can be seen in panels (d) and (e). The symmetry of the interference pattern is well-understood since the scattering event affects only the anti-symmetric coordinate $x=\frac{x_1-x_2}{\sqrt{2}}$. In the last stage of the gate, in panels (f) and (g), the wave-packets return to their initial size and location. The two-particle distribution at the final time is clearly non-separable. To completely characterize the quantum state at the end of the gate, we plot in panel (h) the phase of the wave-function. As expected for a $\sqrt{\text{SWAP}}$ gate, there is a relative phase of $\pi/2$ between the two wave-packets located at $x_1=-1.1645\mu$m and $x_2=1.1645\mu$m and $x_1=1.1645\mu$m and $x_2=-1.1645\mu$m. 

To quantify the performance of the gate, we compute the lower bound on the gate fidelity over all possible initial spin super-positions. Mathematically, the operation we actually do is denoted by the propagator $U({{T}_{\text{gate}}},0)$, and the fidelity is then defined as ${\mathcal{F}=\underset{s,s'\in \{\uparrow ,\downarrow \}}{\mathop{\min }}\,\left|\langle {{\varphi }_{ss'}}|U_{\sqrt{SWAP}}^{\dagger }U({t_{\text{gate}}},0)|{{\varphi }_{ss'}}\rangle\right|^2}$, where ${\varphi }_{ss'}$ is the initial state of the two atoms in the tweezers and $U_{\sqrt{SWAP}}$ is the propagator of an ideal $\sqrt{\text{SWAP}}$ gate. With this definition, we obtain a gate fidelity of $\mathcal{F}=0.9979$. This fidelity is obtained when the spin states are pointing in opposite directions (i.e., $s\ne s'$). For parallel spins, the fidelity is even higher.

\begin{figure}[ht]
	\centering
	\includegraphics[width=1.0\linewidth]{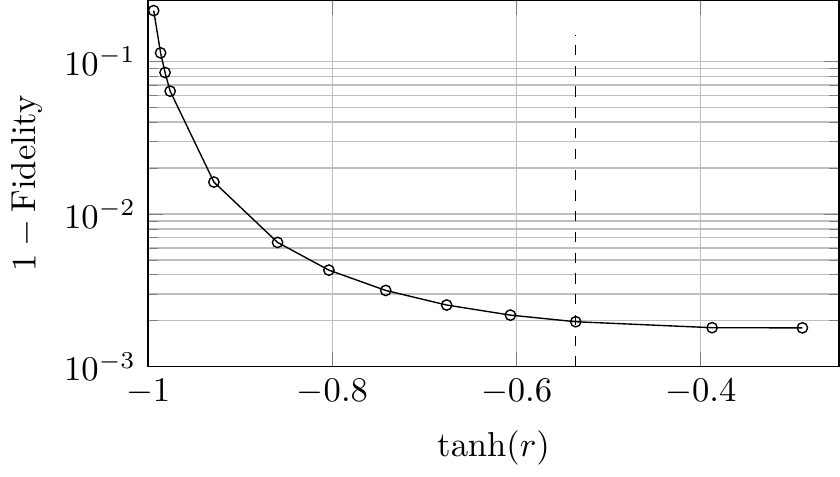}
	\caption{\textbf{The improvement in the fidelity as the squeezing parameter, $\boldsymbol{\tanh(r)}$, increases.} The squeezing parameter is varied by changing the initial width of the atomic wave-packets, in accordance with Eq.(\ref{Eq_squeezing_parameter}). The central harmonic potential is unchanged in these calculations. As $\tanh(r)$ increases, the fidelity converges rapidly towards unity. The dashed line marks the working conditions of the calculation shown in Fig. \ref{fig:two_particle_simulation}.}
	\label{fig:Fidelity_vs_squeezing}
\end{figure}

We now investigate numerically the convergence of the fidelity to unity as $\tanh(r)\rightarrow 1$. To this end, we repeat the simulations and vary the initial width of the atomic wave-packets - taking as initial conditions Gaussian wave-packets with increased (or lowered) widths matching increased (or lowered) squeezing parameters $\tanh(r)$. According to Eq.(\ref{Eq_squeezing_parameter}), this corresponds to changing the squeezing parameters. In each calculation, we optimize the interaction to yield the highest fidelity, namely we scan the prefactor of the $\sin$ function in Eq.(\ref{Eq_gamma_t_2}). The fidelity versus $\tanh(r)$ is shown in Fig. \ref{fig:Fidelity_vs_squeezing}. Its improvement as $\tanh(r)$ increases is very rapid. We also mark the conditions of the calculation done in Fig. \ref{fig:two_particle_simulation} as a dashed line. Note that as $\tanh(r)$ increases, the width of the wave-packets at the collision time becomes smaller. In order to achieve the required numerical accuracy, it is necessary to increase substantially the number of spatial divisions. This limits the maximal value of $\tanh(r)$ we can simulate to around $-0.3$, where the fidelity reaches $\mathcal{F}\approx0.998$.  

\section{A fast gate with scale-invariant driving}\label{sec_STA}
In this section, we generalize our fast gate scheme to include a time varying harmonic trap. There are two main advantage to this extension: Firstly, it allows continuous time dependent control for the harmonic trap, as required in every realistic implementation. Secondly, the gradual increase of harmonic trap depth improves the fidelity of the gate since initially it widens the atoms' wave-packets, hence effectively increasing the squeezing parameter. 

Since our ultimate goal is to complete the gate in a short duration, there is a risk that fast non-adiabatic changes in the harmonic confinement will lead to unwanted excitations that will eventually harm the gate fidelity. To avoid this problem, we adopt a scale-invariant driving strategy which is a well-known technique in the field of shortcut to adiabaticity (STA) \cite{lewis1969exact,lewis1982direct,dhara1984feynman,deffner2014classical}. 

Lewis and Riesenfeld noted that the solutions of the Schr\"odinger equation for a time-dependent Hamiltonian can be written as superpositions of eigenstates of a dynamical invariant \cite{lewis1969exact}. Dhara and Lawande \cite{dhara1984feynman}, and Lewis and Leach \cite{lewis1982direct}, showed that for Hamiltonians of the form,
\begin{equation}\label{HscalingHamonic}
H(t)=
-\frac{{{\hbar }^{2}}}{2m}{{\nabla }^{2}}  +\frac{m}{2}\omega(t)^2\mathbf{x}^2+\theta {{(t)}^{-2}}V\left( \theta {{(t)}^{-1}}\mathbf{x} \right),
\end{equation}
where $\theta(t)$ is a time-dependent scaling factor, and $V(\mathbf{x})$ is an arbitrary potential, have a quadratic-in-momentum invariant,
\begin{equation}
\begin{split}\label{I2}
I_{2}(t)=  \frac{{{\hbar }^{2}}}{2m}&{\left[-i\theta(t)\nabla-m\dot{\theta}(t)\mathbf{x} \right]^{2}} + \\ & +\frac{1}{2}k_0\mathbf{x}^2+\theta {{(t)}^{-2}}V\left( \theta {{(t)}^{-1}}\mathbf{x} \right),
\end{split}
\end{equation}
for some constant value $k_0$, provided that $\omega(t)^2$ and $\theta(t)$ satisfy the Ermakov condition,
\begin{equation}\label{Ermakov}
\ddot{\theta}(t)+\omega(t)^2\theta(t)=\frac{k_0}{m\theta(t)^3}.\end{equation}
Any wave-function $\psi(t)$ which solves the Schr\"odinger equation with $H(t)$ can be written in terms of eigenvectors $\psi_n$ of the invariant (\ref{I2}),
\begin{equation}\label{eq:psi_STA_I}
\psi(\mathbf{x},t)=\sum_n{c_n e^{i\xi_n(t)}\psi_n(\mathbf{x},t)},
\end{equation}
where $c_n$ are constant coefficients, and $\xi_n$ are the Lewis-Riesenfeld phases, given by ${{{\xi }_{n}}(t)=\frac{1}{\hbar }\int_{0}^{t}dt'\langle {{\psi }_{n}}({t}')|i\hbar \frac{\partial }{\partial t}-H(t')|{{\psi }_{n}}(t')\rangle }$ \cite{lewis1969exact}. Importantly, $\psi_n(t)$ have the form \cite{dhara1984feynman},
\begin{equation}\label{eq:psi_STA_II}
\psi_n(\mathbf{x},t)=e^{i\frac{m\dot{\theta}(t)\mathbf{x}^2}{2\hbar\theta(t)}}{\theta(t)^{-D/2} }\phi_n\left(\theta(t)^{-1}\mathbf{x}\right)
\end{equation}
where $D$ is the spatial dimension and $\phi_n$ is the solution of the stationary Schr\"odinger equation at $t=0$ with an eigenvalue $E_n$. The Lewis-Riesenfeld phases are then calculated to be ${\xi_n(t)=-\frac{{{E}_{n}}}{\hbar }\int\limits_{0}^{t}{\frac{1}{\theta {{(s)}^{2}}}ds}}$.

It follows that, for every solution $\phi (x,t)$ of the Schr\"odinger equation with the stationary Hamiltonian, 
\begin{equation}
     H_0=-\frac{{{\hbar }^{2}}}{2m}\frac{{{\partial }^{2}}}{\partial {{x}^{2}}}+\frac{1}{2}{{k}_{0}}{{x}^{2}}+V(x) \ \ ,
\end{equation}
we can find a solution $\psi (x,t)$ for the time-dependent Hamiltonian (\ref{HscalingHamonic}) that has the form,
\begin{equation}\label{Eq_general_psi_for_scale_invariant_driving}
    \psi (x,t)=\frac{1}{\theta {{(t)}^{D/2}}}\exp \left( im\frac{\dot{\theta }(t){{x}^{2}}}{2\hbar\theta (t)} \right)\phi \left( \theta {{(t)}^{-1}}x,\tau (t) \right) \ \ ,
\end{equation}
with ${\tau (t)\equiv \int\limits_{0}^{t}{\frac{1}{\theta {{(s)}^{2}}}ds}}$. We see, therefore, that the solutions with a scale-invariant driving are related to the solutions of the stationary Hamiltonian by a rescaling of the time, an additional position-dependent phase, and a normalization factor. Note that in the stationary limit, $\theta(t)=1$, the rescaled time is identical to the regular time, $\tau(t)=t$, and the solution of Eq.(\ref{Eq_general_psi_for_scale_invariant_driving}) reduces to the solution of the stationary Hamiltonian.

We now apply these results to our case. As explained before, the problem naturally decomposes into single-particle equations for the coordinates $x$ and $X$, given in Eqs.(\ref{harmonic})-(\ref{contact_eq}). We consider first the non-interacting case ($\gamma=0$), in which case the dynamics for both coordinates is the same. For our gate, we only need the central harmonic confinement, hence we set ${V=0}$. Combining the results of Eqs. (\ref{wave_packet_expr}) and (\ref{Eq_general_psi_for_scale_invariant_driving}), we obtain the solution to the one-dimensional time-dependent Schr\"odinger equation with the Hamiltonian $H(t)=-\frac{{{\hbar }^{2}}}{2m}{{\nabla }^{2}}  +\frac{m}{2}\omega(t)^2\mathbf{x}^2$,
\begin{equation}\label{Eq_squeezed_state_sol_for_scale_inv_driving}
\begin{split}
&{{\tilde{\varphi }}_{\pm }}(x,t)=\frac{1}{\theta {{(t)}^{1/2}}}e^{ \left( im\frac{\dot{\theta }(t){{x}^{2}}}{2\hbar\theta (t)} \right)}{{\left( \frac{A(\tau (t))}{\sqrt{\pi }} \right)}^{1/2}}\times\\
&\times{{e}^{-\frac{i}{2}\Theta (\tau (t))}}{{e}^{-\frac{{{\left| \theta {{(t)}^{-1}}x\pm {{x}_{c}}(\tau (t)) \right|}^{2}}}{w{{(\tau (t))}^{2}}}}}{{e}^{\pm i{{p}_{c}}(\tau (t))\theta {{(t)}^{-1}}x/\hbar }} \ \ .
\end{split}
\end{equation}
Notice that while $\theta(t)$ does not appear explicitly in the Hamiltonian, it is related to $\omega(t)$ through the Ermakov equation (\ref{Ermakov}). Importantly, since the solution of Eq.(\ref{Eq_squeezed_state_sol_for_scale_inv_driving}) has to identify with the ground state of the tweezers at the beginning and ending of the gate, the position-dependent phase has to be zero at these times. This requires $\dot{\theta}(0)=\dot{\theta}(t_\text{gate})=0$. In addition, without interactions the gate should swap the positions of the atoms, hence we require  $x_c[\tau(t_\text{gate})]=-\frac{d}{\sqrt{2}}\Rightarrow \int\limits_{0}^{t_\text{gate}}{\frac{1}{\theta {{(s)}^{2}}}ds}=\frac{\pi}{\omega_0}$.

Next, we include the interaction between the atoms, which affects the $x$ coordinate. The Hamiltonian reads 
\begin{equation}\label{HscalingHamonic_with_interactions}
H(t)= -\frac{{{\hbar }^{2}}}{2m}\frac{{{\partial }^{2}}}{\partial {{x}^{2}}}+\frac{m}{2}\omega (t)^2 x^2+\tilde{\gamma }(t)\delta (x) \, \, .
\end{equation}
Similar to what we have done in section \ref{sec_fast_gate}, we treat the problem as a scattering process of $\tilde{\varphi}_+ (x,t)$ from a delta-potential located at $x=0$, with transmission and reflection coefficients, $T$ and $R$. We impose the same boundary conditions of Eqs.(\ref{Eq_boundary_condition_1})-(\ref{Eq_boundary_condition_2}) and demand $R/T=\pm i$ to realize a perfect $\sqrt{\text{SWAP}}$. This yields
\begin{equation}
\begin{split}
\tilde{\gamma }(t)=&\mp i\theta {{(t)}^{-1}}\frac{{{\hbar }^{2}}}{m}\left[ -\frac{i}{\hbar }\frac{m\omega_0 d}{\sqrt{2}}\sin (\omega_0 \tau (t))+\right.\\
&\left. -\frac{m\omega_0 }{\hbar }\left[ \frac{1-\tanh (r){{e}^{-2i\omega_0 \tau (t)}}}{1+\tanh (r){{e}^{-2i\omega_0 \tau (t)}}} \right]\frac{d}{\sqrt{2}}\cos (\omega_0 \tau (t)) \right] \ \ ,
\end{split}
\end{equation} 
which becomes in the limit $\tanh(r)\rightarrow 1$,
\begin{equation}\label{Eq_gamma_t_3}
    \tilde{\gamma }(t)=\mp \sqrt{2}\theta {{(t)}^{-1}}\omega_0 \hbar d\sin (\omega_0 \tau (t)) \ \ .
\end{equation}
This result is a generalization of Eq.(\ref{Eq_gamma_t_2}) and reduces to it in the stationary case. 

To realize the STA driving, the trajectory of the rescaling parameter, $\theta(t)$, has to be smooth and satisfy several conditions,
\begin{enumerate}
\item $\dot{\theta }(0)=\dot{\theta }({{t}_{\text{gate}}})=0$ -- This cancels the position-dependent phase in Eq.(\ref{Eq_squeezed_state_sol_for_scale_inv_driving}).
\item $\omega {{(0)}^{2}}=\omega {{({{t}_{\text{gate}}})}^{2}}=0$ -- The harmonic trap vanishes at the beginning and ending of the gate.
\item ${{\omega }_{0}}\int\limits_{0}^{{{t}_{\text{gate}}}}{\frac{1}{\theta {{(s)}^{2}}}ds}=\pi $ -- This ensures the SWAP condition with no interactions. 
\item $\omega {{(t)}^{2}}\ge 0$. This condition is required from a practical perspective, to avoid the need to change from attractive to repulsive harmonic potential. A repulsive potential can be generated using an optical trap at a blue-detuned wavelength \cite{Grimm2000}. Having both repulsive and attractive potentials during the gate's operation requires changing the wavelength of the optical trap, which is better to avoid. Therefore, we seek for a driving that keeps the harmonic trap purely attractive. Note that this condition translates into a condition on $\theta(t)$ through the Ermakov relation, $\omega {{(t)}^{2}}=\frac{1}{\theta (t)}\left( \frac{k_{0}^{{}}}{m\theta {{(t)}^{3}}}-\ddot{\theta }(t) \right)$.
\end{enumerate}

There are infinitely many choices of driving that will satisfy the conditions above. We adopt the following parametrization,
\begin{equation}\label{Eq_theta_t}
    \theta (\tilde{t})={1+{{a}_{2}}{{\tilde{t}}^{2}}+b_1{{\tilde{t}}^{{{\beta }_{1}}}}+b_2{{\tilde{t}}^{{{\beta }_{3}}}}}.
\end{equation}
with $\tilde{t}\equiv\left| \frac{2t-t_\text{gate}}{{{t}_{\text{gate}}}} \right|$. Conditions 1-2 (3) yield linear (non-linear) relations on the five parameters ${{\beta }_{1}}$, ${{\beta }_{2}}$, ${{a}_{2}}$, $b_1$ and $b_2$. Condition 4 is an inequality condition, and $\omega {{(t)}^{2}}$ is continuous if we choose $\beta_1,\beta_2>2$. 

\begin{figure}[ht]
\centering
	\includegraphics[width=1\linewidth]{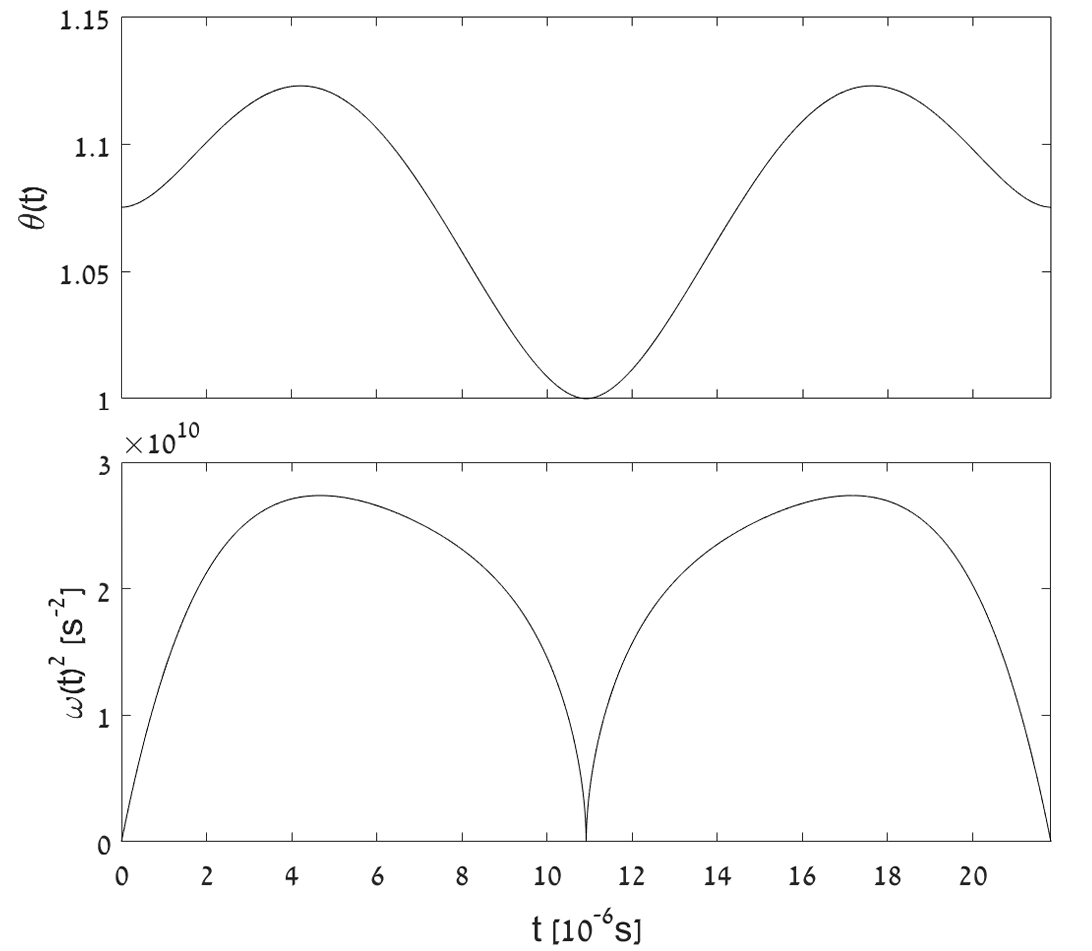}
	\caption{The rescaling function, $\theta(t)$, and the harmonic potential, $\omega(t)^2$. The STA driving protocol of $\theta(t)$ is defined in Eq.(\ref{Eq_theta_t}), with ${{\beta }_{1}}=2.548$, ${{\beta }_{2}}=6.227$, ${{a}_{2}}=1.684$, $b_1=-1.808$ and $b_2=0.199$, and $t_\text{gate}=21.84\mu\text{s}$. The Ermakov equation (\ref{Ermakov}) gives $\omega(t)$, with $k_0=m\omega_0^2$ that is set by the third condition, ${{\omega }_{0}}\int\limits_{0}^{{{t}_{\text{gate}}}}{\frac{1}{\theta {{(s)}^{2}}}ds}=\pi $.}	\label{fig:theta_and_omega}
\end{figure}

We employ a numerical optimization code to find parameters that satisfy these conditions and yield optimal performance for the gate. In Fig. \ref{fig:theta_and_omega} we plot an example of such driving protocol, with ${{\beta }_{1}}=2.548$, ${{\beta }_{2}}=6.227$, ${{a}_{2}}=1.684$, $b_1=-1.808$ and $b_2=0.199$ which were found for the same parameters as those used in the stationary gate of Fig. \ref{fig:fast_gate_operation}, namely $t_\text{gate}\approx 21.84\mu\text{s}$, a central Gaussian potential, which approximates the harmonic one, with a waist of $\sigma=11.857\mu$m. In this example, the maximum depth is $\max[U(t)] =694\mu \text{K}\times k_B$, only moderately higher than the value used in the stationary gate with the same duration. With these parameters, the fidelity of the scaled invariant gate is $f=0.9988$. For comparison, an equivalent gate with a constant Gaussian potential depth equal to $\max[U(t)]$ gives a gate time of $18.993\mu$s and a fidelity $f=0.9979$. The differences between these numbers are not significant given our numerical accuracy. 

\section{Discussion}\label{sec_discussion}
In this work, we have presented a new concept for a universal $\sqrt{\text{SWAP}}$ gate performed on two fermionic atoms trapped in optical tweezers. The gate is based on releasing the two atoms in a central harmonic trap and exploiting the phase accumulated during the scattering process. The big advantage of this scheme is that the gate-duration is set by the harmonic trap period, and thus can be very short -- on the order of $10\mu$s. Moreover, the gate duration is independent of the initial distance between the atoms. Tuning the gate for optimal performance is simple, and done by tuning the interaction strength between the atoms (e.g., working near a Feshbach resonance). We have studied numerically the performance of the gate at experimentally realistic parameters, and have shown that the gate is robust and achieves very high fidelity above 0.998.

We have also given a generalized version of the gate when the harmonic confinement is time-dependent. Using scale-invariant driving, we have derived a general form for the harmonic frequency, $\omega(t)$, which ensures that the fidelity of the gate is not compromised due to the non-adiabatic changes. In fact, this time-dependent scheme improves the fidelity as it leads to larger squeezing of the colliding atomic wave-packets. It also allows for a gradual turning on and off of the harmonic potential -- a plausible practical advantage.

We expect that the ability to implement a robust and high-fidelity universal gate at such short timescales will push forward the field of neutral atoms in optical tweezers. This platform offers huge advantages in terms of scalability and controllability, but historically suffered from relatively slow and low fidelity of the two-qubit gates. Our approach accelerates dramatically the attainable rate of quantum computation in this platform, and also adds the flexibility to perform two-qubit gates between qubits at different distances. Thus, all-to-all connectivity, which greatly strength the computational power of a given number of qubits, can be achieved. 

\begin{acknowledgments}
We thank Gal Ness for helpful comments on the manuscript. This research was supported by the Israel Science Foundation (ISF), grants No. 1779/19 and 218/19, by the United States - Israel Binational Science Foundation (BSF), grant No. 2018264, and by the Pazy Research Foundation.
\end{acknowledgments}
	
%\bibliography{main_v1.bib}


\end{document}